# Sedimenting microrollers navigate saturated porous media

Samuel R. Wilson-Whitford[1,2]\*, David Kramer[1], Jinghui Gao[1], Maria Chiara Roffin[3] and James F. Gilchrist[1,2]\*

[1]Department of Chemical and Biomolecular Engineering, Lehigh University, Bethlehem, Pennsylvania, USA, 18015

[2]Aston Institute for Membrane Excellence, Aston University, Birmingham, UK, B4 7ET.

[3]School of Physics and Astronomy, University of Leicester, Leicester, UK, LE1 7RH.

Corresponding Email\*: s.wilson-whitford@aston.ac.uk

Corresponding Email\*: gilchrist@lehigh.edu

## Abstract

Particle sedimentation through porous media is limited by the inability of passive material to overcome surface interactions and a tortuous network of pores. This limits transport, delivery, and effectiveness of chemicals used as reactants, nutrients, pesticides, or for waste remediation. This work develops magnetically responsive microrollers that navigate the complex interstitial network of porous matter. Rather than arresting on the upward facing surfaces of the pores, particles can roll and fall further, increasing transport by orders of magnitude. This work directly investigates Janus microrollers, activated by a rotating magnetic field, rolling and sedimenting though an index-matched porous medium. The mechanism of enhanced transport is determined, and the material flux is primarily a function of microroller concentration, rotation rate, and magnetic field strength. This mechanism is most efficient using a minimum number of rotations spaced out periodically in time to reduce the required energy input to greatly enhance transport. This general mechanism of transport enhancement can be broadly applied in numerous applications because the particles delivered within the porous matrix may be comprised of a wide variety of functional materials.

## Introduction

Sedimentation-driven transport of granular media is ubiquitous in nature and industrial powder-handling processes.[1–4] In systems where the size asymmetry of particles is large, smaller particles can percolate through the interstitial regions of larger particle assemblies saturated with a fluid.[5–7] At lower grain size ratios, $R = d_l/d_f < 3$, where $d_l$ is the diameter of the larger grains and $d_f$ is the diameter of the smaller grain,[8] shear-enhanced percolation segregation of granular media, via gravity driven flow[9–11] or vibration,[12–14] results in a strongly partitioned system with smaller particles concentrating near the bottom of the flow. However, at more significant size ratios, $R > 3$, in what is considered the "fine particle" regime, such as in porous materials and jammed systems where the larger grains are nearly motionless, partitioning by flow and/or gravity occurs by filtration[8,15,16] or alternatively by adsorption when an affinity between the small and large particles exists.[17]

In the simplest case, when there is no gradient in pressure on the fluid within the porous media, the description of percolation and filtration by sedimentation remains relatively complex. Smaller constituents, whether colloidal or inertially-dominated in scale, will settle until geometrically constrained and/or adsorbed to an interface. On the pore-scale, phenomenon such as electrostatic and van der Waals *particle-particle* and *particle-media* interactions can have a large effect on pore clogging and particle deposition, especially on the colloidal scale.[17–19] In addition to flows of passive colloids, there are also an increasing number of studies looking at the movement of responsive or active colloids through complex periodic geometries.[20–22] For granular scale particles, transport through a porous structure is dominated by sedimentation rather than Darcy flow through porous media. In a porous network of discrete non-cohesive grains, where particles size ratio is very large, $R \gg 3$, particles falling onto the top plane of the saturated porous structure will come to rest on the surface of grains, providing that the angle tangential to the point of particle-grain contact is sufficiently shallow, and the contact made arrests the particle through adhesive forces or friction. If the angle is steeper, particle inertia or lubrication will overcome the adhesion to the grain and the particle will fall though to interact with interfaces deeper within the network. At higher volume fractions of particles, or at lower $R$, there are additional considerations such as jamming and classic granular scale filtration problems.[23,24] Overall, a combination of factors, such as size, aspect ratio, Reynold number, friction etc. can be considered in a general assessment of sedimentation and flow of granular scale materials through porous materials, as described by Harris.[25,26]

The tortuous nature of porous materials is what makes them both widely useful and fundamentally problematic. They are extremely effective for filtration, scaffolds, absorption, and sensing but at the same time can be inhibitive to material transport and recovery, such as oil, microplastic and ground water pollutant remediation.[27–30] Whether looking at the positive or negative aspects of the materials, the ability to control the movement of matter through porous media is of critical importance to areas such as agriculture, mining, and water and waste management.

This work aims to design systems of externally manipulated microrollers that can navigate porous media through enhanced percolation sedimentation. To do so, it is necessary to develop an understanding of how microrollers behave in porous media. Spherical microrollers that experience torque in response to a rotating magnetic field, similar to those employed in confined microrobotics,[31,32] are used to overcome particle arrest during high $R$ sedimentation filtration, as shown in Fig. 1 and demonstrated in V1. When a suspension of fine, magnetically-responsive granular particles are introduced to a saturated porous network of static beads, after a short time, without particle rotation, all particles in a porous network are immobilized on the surface of beads (Fig. 1a,c-d). Subsequent application of torque through an external motorized rotating magnetic field remobilizes these trapped particles (Fig. 1b,e-f), thus reinitiating the sedimentation process and preventing immobilization due to the particles' rotation and resulting lubrication interaction with the pore interfaces, V1. This approach, characterized herein, allows for control of enhanced gravity driven transport of particulate matter in the absence of an applied pressure and flow.

# Experimental

## Janus Particle Synthesis

Janus particle preparation was identical to that described in a previous publication.[33] In summary Sub-monolayer films of 38-53 µm (measured; d[3,2] 43.6 ± 5.9 µm) diameter PMMA particles were prepared on a roll-to-roll scale by automated Langmuir-Blodgett.[34] The dried films were coated with 100 nm of Fe by physical vapor deposition. Janus particles were removed from the substrate by sonication in ethanol and then air dried.

## Experimental set-up

See Supplementary Information for full diagram of experimental set-up. 300 gel beads are hydrated in 1 M NaCl solution to reduce maximum swelling diameter. The 300 swollen beads are added to a clear plastic container of dimensions 40 mm x 40 mm x 55 mm (W x D x H) and covered with more NaCl solution until the fluid level is at least 1 cm higher than the top layer of beads. Janus particles are washed by cycling ethanol washes and magnetic separation to remove any uncoated particles or dust. After washing and drying, the desired mass of Janus particles is added to the sample container. The sample is shaken gently to disperse the particles homogenously in order to take a measurement at $t = 0$. The container is positioned in a holder to give reproducible sample positioning. The sample container is back-lit with and LED lighting panel.

The rotating magnet and measured field strength data used in this publication are reproduced from a previous publication.[33] 4 x 1000 Gauss neodymium magnets were attached to a motorized wheel at 90 º intervals. The outward facing pole was alternated, 0° = N, 90° = S, 180° = N, 270° = S. Rotation of the magnetic field is clockwise relative to the camera view during data acquisition. The field strength relative to orientation was measured experimentally using a gaussmeter. The rotating magnet is positioned above the sample container using a frame which has adjustable height, so that the distance between the magnet and sample can be adjusted. Data was collected using a Nikon DSLR camera.

## Data Analysis

For large particles casting shadows, $c \propto I^{-1}$. Videos of runs were loaded into MATLAB using the VideoReader package. Videos are converted to 8-bit and inverted (black background, white particles). The max white and black values are selected by averaging of pixel intensities in specific regions of the video. The analysis ROI is set and the average pixel intensity over the sample is calculated for each frame of the 24 fps videos. A background intensity, taken from a pixel intensity average at the top of the sample in the final frame, is subtracted from every frame. The inherent periodicity in the intensity of the rotating Janus particles is averaged out through the use of a 23-point moving average.[35–37] The averaged intensity vs time data is then reduced to 1 fps. The peak intensity after magnetic rotation was initiated, was set as $t = 0$ and considered $c_{max}$ for purposes of normalisation.

## Results and Discussion

A suspension of magneto-responsive Janus microrollers (~44 µm, Fig. S1-2) was added to a porous media consisting of hydrogel beads (5.4 ± 0.4 mm when swollen in 1 M NaCl), $R = d_{bead}/d_{particle} \sim 113$.[8] The beads and fluid are transparent nearly optically index-matched and back-lit, allowing visualization of Janus particle transport and sedimentation through the 3D network.[38] The particles themselves are not transparent, and the concentration of particles is estimated from the light intensity passing through the index-matched medium in a region several particles deep (Fig. S3). This is quantitative when particles are mobilised and sedimenting, but apparent concentrations are qualitative when the system is in a state of full rest, i.e. arrested particles.

In existing microrobotic studies, microrollers have been used to traverse challenging geometries or confinements through continuous driving of isolated granular scale microrollers, even under opposing flows.[31,32] Similarly, the collective motions of dense suspensions or beds of granular microrollers has previously revealed surprising physics, demonstrating that dense piles of these Janus particles exhibit uphill heaping when activated by a rotating magnetic field where the particles rotate roughly in phase with the magnet and the effective interparticle friction is a function of the magnetic field strength.[33] Strong magnetic fields lead to the formation of transient particle chains, rather than permanent chains formed by their micron-scale counterparts.[35,39] Here, designing a system to navigate a complex porous network, the same ferromagnetic Janus particles are studied in dilute/semi-dilute conditions, $\phi_p$ < 0.02, where their sedimentation dynamics should be comparable to that of spherical and rod-shaped particles,[40,41] assuming that a proportion of particles have formed short chains. This observed in V1 and V2.

This work is in the limit where percolating particles are significantly smaller than the interstitial spaces between the porous network, ($R \gg R_t$ where $R_t = d_{bead}/d_{pore} \sim 6.5$ and $R \sim 113$) where no pore clogging or jamming is observed.[8] Microroller suspensions are added to the porous medium such that the total volume fractions with respect to the continuous interstitial volume, is up to $\phi_p$ = 1.5×10⁻². The system is agitated to generate a relatively homogenous distribution of particles throughout the network (Fig 1c). Particles first freely sediment until they come to rest in the interstitial sites of the structure through gravitational sedimentation (Fig 1a,c-d) as a result of coming into frictional contact with the upward facing surfaces of the porous gel bead matrix. This particle distribution is static until rotational torque is applied by rotating magnets positioned above the sample, parallel to the plane of observation and normal to the container sides. This applied rotating magnetic field leads to particle redispersion and gravity driven sedimentation through the network until all particles have either passed through the porous medium or come to rest (Fig. 1b,e-f). The size of the Janus particles and the mass of the iron oxide deposition is large enough to provide sufficient magnetic torque to overcome the barrier to particle desorption from the bead-particle interface. This study is limited to a single particle size, but will be applicable to any system in which the bead- particle ratio, $R \gg 3$, and $d_{particle}$ is non-Brownian. At the highest concentration, the downward trajectory of particles also instigates upward streamers of some particles due to fluid displacement.[42,43] The deliberate positioning of the magnet above the sample removes any concern that the particles are being magnetically pulled downward through the porous network. For these experiments, there exists a separation distance between the sample and the magnet, $h_{max}$, at which no appreciable particle agitation and subsequent transport occurs. Distance, $h$, is measured from the bottom of the packing of beads (Fig. S4-S5). The magnet field, $\beta$, felt by particles at the bottom of the sample at $h_{max}$, is defined as $\beta_0$. Therefore, each experiment can be defined as having a dimensionless magnetic field strength, $\beta/\beta_0$, where $\beta \propto h^{-2}$, as mentioned in our previous work.[33] A fit of the experimentally measured field is given in the Electronic Supporting Information (Fig. S4-S5). The particles either move as individual microrollers or short chains of particles that rotate in response to the rotating field, V2. Gel bead packing is roughly hexagonally close-packed such that sedimenting particles interact with beads in layers below and there are no vertical pores that allow transport straight to the bottom of the container. It should be noted, that during image analysis, the bottom 2.5 mm of the sample is not analysed to allow particles to leave the region of interest as they are transported through the network. What follows is quantification of the dynamics of this system as a function of rotation rate, $W$, magnetic field strength, and initial concentration. This is the type of information one needs to engineer a system that takes advantage of this mode of enhanced transport. The analysis of the concentration dependence is taken over a near-full sample region of interest to avoid the inconsistencies that result from the spatial heterogeneity caused by local concentration fluctuations as particles fall onto the surfaces of the porous media.

**Influence of field strength, rotation rate and volume fraction**

High-definition recordings of torque-initiated sedimentation were analysed to show relative intensity changing over time. The evolution of the normalized intensity, correlated with concentration, is studied for varying applied magnetic field, $(\beta/\beta_0)$, rotation rates, $\Omega$, and particle volume fraction relative to total interstitial volume, $\phi$, as shown in Fig. 2. Each of these experiments, without any particle rotation, would remain at a maximum apparent concentration, $c/c_{max} = 1$, indefinitely. As torque is applied, particles detach and roll off the top surfaces of the porous media (V1) and subsequently fall through the interstitial spaces. Decrease in apparent concentration follows what initially appears to be a pseudo-exponential decay, starting at $c/c_{max} = 1$ and decaying to a plateau, i.e. when no further particles are filtered from the system. The first and most obvious observation is that the rotation rate of particles does not influence transport, however volume fraction of particles and magnetic field strength strongly influence transport. Note the rotating magnet is positioned above the sample and the primary force through the network is gravity assisted by imposed rolling.

A clearer picture of the influence of volume fraction and field strength can be seen by breaking down Fig. 2 into 3 separate log-lin plots based on the volume fraction of particles (Fig. 3 a-c). For each plot, it is yet clearer that the filtration of particles is independent of particle rotation rate. In each data set, close to time zero, the filtration flux $-dc/dt$, is shallower due to initial agitation and redispersion of particles but is then followed by a steep decline in the apparent concentration of particles in the system. At longer timescales, this tends towards a plateau, at which point the maximum amount of particles have been filtered through the system, leaving behind a fraction of particles trapped in the network. For each volume fraction, successive increases in the applied field increase the fraction of particles successfully filtered through the porous network. Fig. 3 tells us that while one might assume that the speed at which a particle rolls across the surface would strongly influence particle motion, torque is primarily acting as the activation energy to overcome the particle adhesion to the bead surface. After this point, particle rotation simply prevents reabsorption to bead surfaces through localized particle lubrication and transport is primarily facilitated by sedimentation.[44]

Similarly, comparing samples agitated at the same applied field strength across the increasing volume fraction reveals a similar relative increase in the fraction of particles filtered through the network i.e. at longer timescales the plateau of the relative concentration shown in Fig. 3 is higher for $\phi_p = 1.7 \times 10^{-3}$ as compared to $\phi_p = 3.3 \times 10^{-3}$ and $5.0 \times 10^{-3}$. It is known that the presence of magnetic permanent dipoles in particles, especially Janus particles, can lead to particle chaining,[35,39] with longer chains forming at higher field strengths and volume fractions. Chaining of particles is also observed for this system, as seen in V1 and V2, it is therefore logical to suppose that particle chaining or clustering could lead to sedimentation at a higher velocity once desorbed particles are in lubricated free-fall, due to higher masses and orientation-dependent sedimentation velocities.[45,46] At lower volume fractions, there is a reduced capacity to form chains and therefore there is a reduction in the average sedimentation velocity of particles.

Both the ability to drive particle rotation, and therefore overcome adhesion to beads, and the ability to form chains is a function of the field strength, $(\beta/\beta_0)$. To investigate the influence of field strength more closely, a sample containing $\phi_p = 1.7 \times 10^{-3}$ was subject to a range of relative magnetic field strengths from $1 \leq (\beta/\beta_0) \leq 32$ at a fixed rate of rotation, $\Omega = 3.2$ Hz (Fig. 4). It can be observed clearly that using a stronger relative field strength enhances the rate and total particle filtration in porous systems. Fig. 4b shows the relationship between the relative concentration of particles remaining trapped in the porous system at infinite time, $\psi = c/c_{max}(t_\infty)$, and shows an inverse proportionality $\psi \propto \beta^{-1}$. Since greater field strength provides more applied torque to adsorbed particles, the detachment threshold for particle-bead contacts is more easily achieved at higher relative field strengths, leading to higher particle flux. Additionally, the maximum force of adhesion in the system can be estimated as $F_{ad,max} \propto (\beta/\beta_0)_{\psi=0}$.

**Particulate flux through porous media**

Thus far, analysis of relative concentration has primarily focused on the total amount of material removed from the porous media rather than the rate of Janus particle transport or flux, $Q = -d(c/c_{max})/dt$. The flux, $Q$, is calculated from data in Fig. 2 and plotted in Fig. 5. Following the initial period of agitation, $Q$ decreases pseudo-exponentially towards $t_\infty$. This is initially counterintuitive as if the particles were simply in free-fall and unable to re-adsorb due to rotational lubrication, we would expect $Q$ to be constant and be a function of the terminal velocity of the particles or particle chains. One plausible explanation for the apparent decrease in $Q$ relates to the size distribution of particles and particle chains. A distribution of particle sizes, where longer magnetic chains or aggregates of particles, would give a distribution in sedimentation velocities which would therefore result in a decreasing $Q$ over time.

Fig. 6 shows the maximum flux of filtration, $Q_{max}$, for a system with magnetic field strength and rotation rate fixed at $(\beta/\beta_0) = 5.6$ and $\Omega = 3.2$ Hz, respectively. High volume fractions are directly related to the observed length and count frequency of particle chaining, and as such it would be expected that $Q_{max}$ was dependent on volume fraction. As such the volume fraction of microrollers in the porous network at $t_0$ was varied between for $\phi_p$ = 8.3×10$^{-4}$ – 1.5×10$^{-2}$. The predicted relationship between $Q_{max}$ and $\phi_p$ was observed and is shown to have a linear increase with increasing $\phi_p$ up to $\phi_p \sim 7.0×10^{-3}$. At $\phi_p > 7.0×10^{-3}$ the is a change in behavior which at $\phi_p \to \infty$, would plateau where flux becomes independent of volume fraction. The change in behavior is hypothesised to relate to the volume fraction at which particles and particle chains on beads begin to interact with neighboring particles, providing a cooperative detachment behavior. This behavior would of course become irrelevant once particle concentrations begin to saturate pore throats.

**Energy cost of transport**

It is proposed that this system of microroller enhanced filtration could be used in agricultural, industrial or civil engineering applications where delivery of material through or into porous media is essential. In all such assisted filtration systems, the energy efficiency of the process should be considered. Here, energy consumption is proportional to the total number of rotations at any point in time, $\Delta E \propto N_{\Omega,total}$. As shown in the earlier data, there is a maximum rate at which particles will move through the porous network for a specific set of conditions when rotation is applied continuously, but following detachment/lubrication from the bead surfaces, the particle rotation only serves to prevent re-sedimentation rather than enhancing transport further (Fig. 2 and 3). Therefore, there is a minimum frequency at which particles need to be intermittently agitated to prevent re-sedimentation. In an additional experiment, a sample with $\phi_p$ = 1.7×10$^{-3}$ was agitated intermittently, allowing released particles to come to rest in the interstitial sites and on bead surfaces, followed by re-agitation. The experiment was performed with the intermittent agitation cycles, $\omega$, where the number of particle rotations per cycle, $N_{\Omega,\omega}$ = 2, 10, 20, 30 and where the timescale of an agitation cycle was, $t_\omega$. As such cycle time $t_\omega \propto N_{\Omega,\omega}$. After agitation, the time allowed for particles to re-sediment is fixed at $t_r$ = 60 s in all cases. The results are shown in Fig. 7 and also in the Electronic Supporting Information, where data is aligned without relaxation time (Fig. S6).

As could be predicted, there is a greater time efficiency in utilizing longer rotation cycles. This is due to the sustained lubrication of particles during filtration and the prevention of particle re-adhesion to beads (Fig. 7a). However, an analysis of the change in concentration per cycle, $\Delta c/c_{max}$ normalised by the number of rotations per cycle, $N_{\Omega,\omega}$, i.e. amount of particles filtered per magnetic rotation, it reveals a much higher energy efficiency (Fig. 7b). There is wasted energy in rotating particles which are already in free-fall. For completeness, at $N_{\Omega,\omega}$ = 0, there is no energy input to the system and thus no change in $c/c_{max}$. From Fig. 7c, the total rotations to transport 100% of particles through the porous system can be estimated for all agitation cycle lengths, $N_{\Omega,\omega}$. With a cycle length $N_{\Omega,\omega}$ = 2, the total energy required is reduced by ~75% as compared to $N_{\Omega,\omega}$ = 30. Of course, the exact nature of this behaviour would vary based on porosity, particle size and viscosity amongst other properties.

## Conclusions

Transport through porous media is of great importance to the natural world and to industrial processes with the movement of materials into, through and out of porous materials strongly tied to process economics and sustainability. In this work, the inclusion of granular microrollers to aid and enhance sedimentation driven transport in a complex geometry is a means to counteract irreversible mass stagnation within porous structures. This method not only helps to provide continuous filtration of materials but could also be used to help cooperatively transport passive or active materials through porous structures in a general or targeted way. The fundamental mechanism of microroller transport into porous media is shown to be cooperative torque induced detachment and sustained lubrication. Particle rotation serves only to detach particles and then prevent reattachment to interfaces within the porous media. However, the strength of the applied field and the volume fraction of microrollers is shown to have a significant affect on filtration, with higher field strength and volume fractions increasing the mass flux due to *microroller-microroller* interactions and applied torque during detachment.

We demonstrate that when using microrollers, surprisingly little energy input is needed to significantly enhance the transport of particles through a tortuous medium and as a result has implications in agriculture, through nutrient or seed delivery,[47] materials remediation, and even civil engineering, through targeted delivery of re-enforcing materials. Further specified explorations are necessary for understanding the impact and limitations of particle size and porosity on transport enhancement. The greatest advantages of this proposed technology are its simplicity of execution, easy microroller recoverability, and the ability to implement this technique using a robust set of materials beyond iron-coated PMMA microspheres.

## Author Contributions

SRWW: Conceptualisation, experimental design, experimentation, data analysis, coding, writing and editing, research direction. JFG: Conception, research direction, funding attainment and editing. DJK: Experimentation and writing. JG: Particle synthesis. MCR: Coding and data analysis.

## Conflicts of interest

There are no conflicts to declare.

## Acknowledgements

Funding for this work was provided by Johns Hopkins University, Applied Physics Laboratory. DJK was supported in part by the Lehigh University's Department of Chemical and Biomolecular Engineering Alumni Experiential Learning Award. MCR was funded by NSF grant 1931681.

# Figures

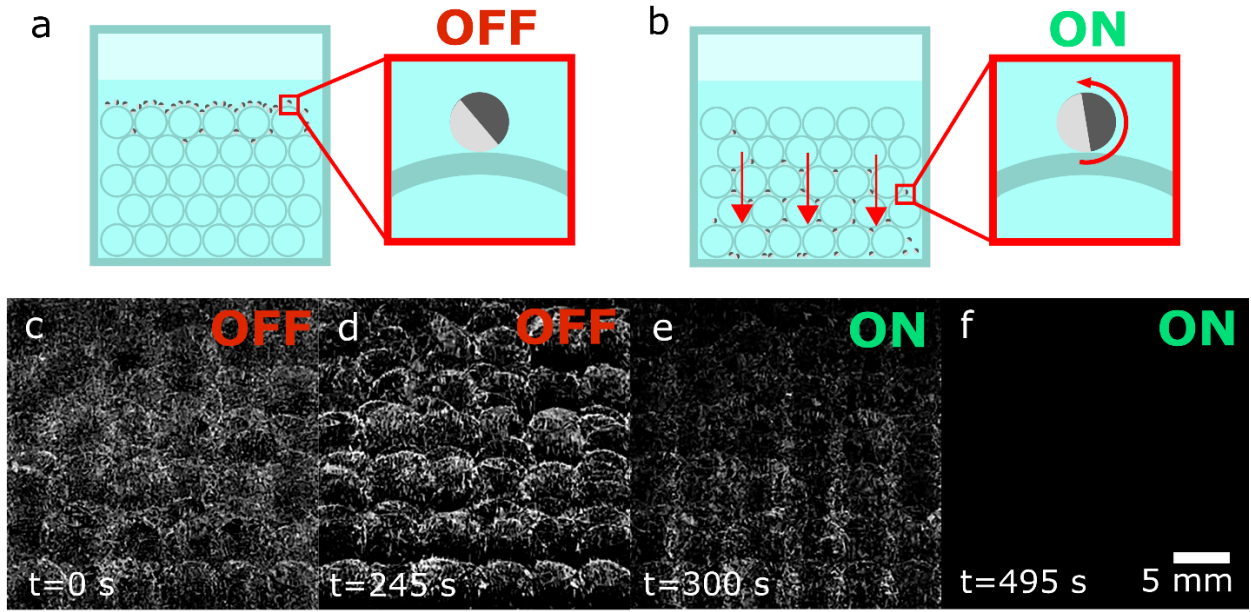

**Fig. 1. Schematic of microroller transport in porous media.** Representation of the experiment (**A**) Janus particles at rest on the surface of beads in a static, saturated porous medium. Rotation $\Omega = 0$ (**B**) Rotating and lubricated Janus particles, detaching and penetrating the porous gel bead network when $\Omega > 0$ (**C**) Subtracted inverted image of Janus particle sedimenting through gravity in network at $t = 0$ s and $\Omega = 0$ (**D**) particles fully sediment in the network at $t = 245$ s ($t/t_s = 1$) and $\Omega = 0$. (**E**) Rotating particles being transported through network at $t = 296$ s ($t/t_s = 1.2$) and $\Omega > 0$ (**F**) All particles transported through the network at $t = 488$ ($t/t_s = 1.99$) and $\Omega > 0$. Images from sample with $(\beta/\beta_0) = 5.66$, $\Omega = 3.2$ Hz, and $\phi_p = 5.0 \times 10^{-3}$.

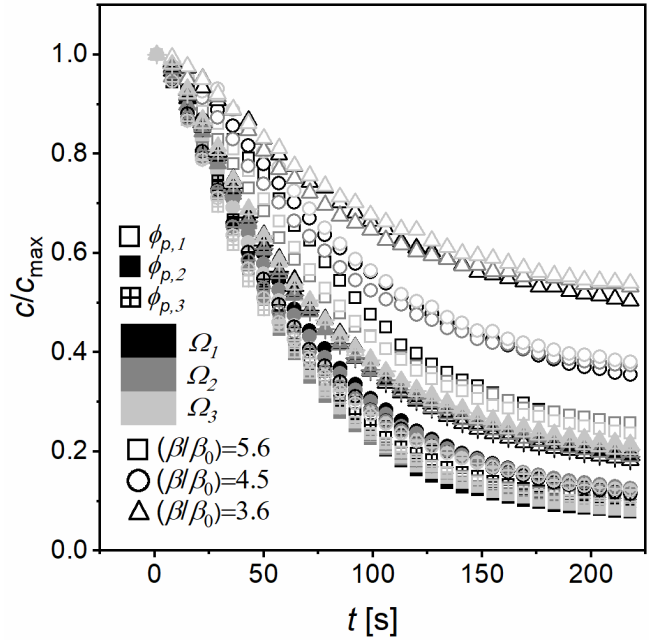

**Fig. 2. Relative concentration vs. time for filtration under magnetic agitation.** Relative concentration vs. time with various initial volume fractions, rotation rates and magnetic field strengths. All data shows enhanced particle removal compared to $\Omega = 0$ when $c/c_{max} = 1$. The symbol legend follows: squares, $(\beta/\beta_0) = 5.6$; circles, $(\beta/\beta_0) = 4.5$; triangles, $(\beta/\beta_0) = 3.6$; black, $\Omega_1 = 1.6\ Hz$; dark gray, $\Omega_2 = 3.2\ Hz$; light gray, $\Omega_3 = 4.8\ Hz$; open, $\phi_{p,1} = 1.7\times10^{-3}$; filled, $\phi_{p,2} = 3.3\times10^{-3}$; and crossed $\phi_{p,3} = 5.0\times10^{-3}$.

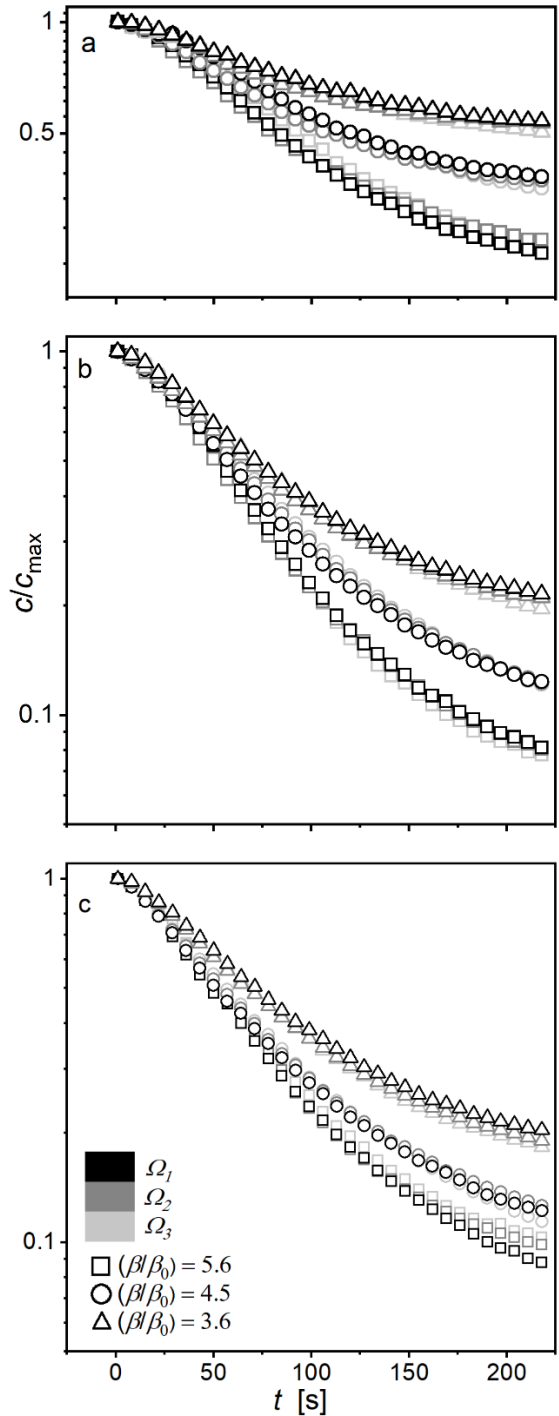

**Fig. 3. Influence of rotation rate, volume fraction and applied field on microroller transport in porous media.** Log-linear concentration change vs time at different $(\beta/\beta_0)$ and rotation rate $\Omega$. The symbol legend follows: square, $(\beta/\beta_0)= 5.6$; circle, $(\beta/\beta_0) = 4.5$; triangle, $(\beta/\beta_0) = 3.6$; black, $\Omega_1 = 1.6\ Hz$; dark gray, $\Omega_2 = 3.2\ Hz$; light gray, $\Omega_3 = 4.8\ Hz$). For each subplot, (**A**) $\phi_p = 1.7 \times 10^{-3}$ (**B**) $3.3 \times 10^{-3}$ and (**C**) $5.0 \times 10^{-3}$.

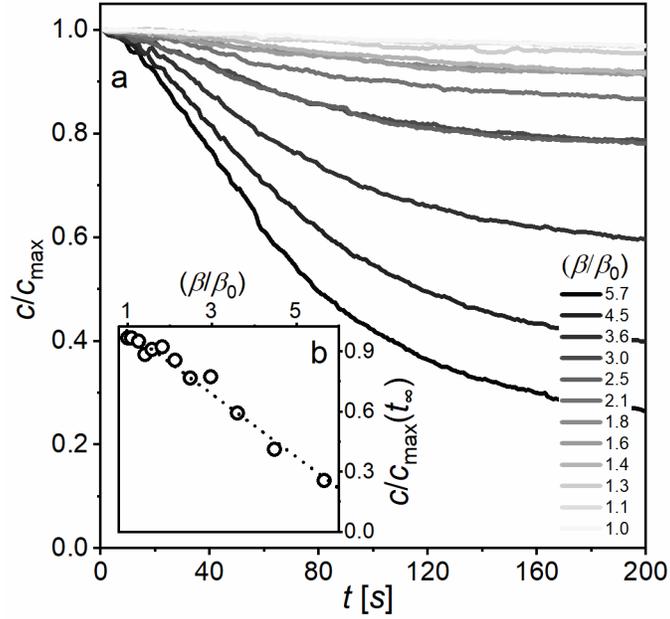

**Fig. 4. Influence of applied field on microroller transport in porous media.** (A) The impact of field strength, $(\beta/\beta_0)$, on the relative concentration for microrollers moving through porous media under the influence of gravity. (B) Inset shows the influence of $(\beta/\beta_0)$ on the relative concentration at infinite time for $\phi_p = 1.7 \times 10^{-3}$ and $\Omega = 3.2\ Hz$.

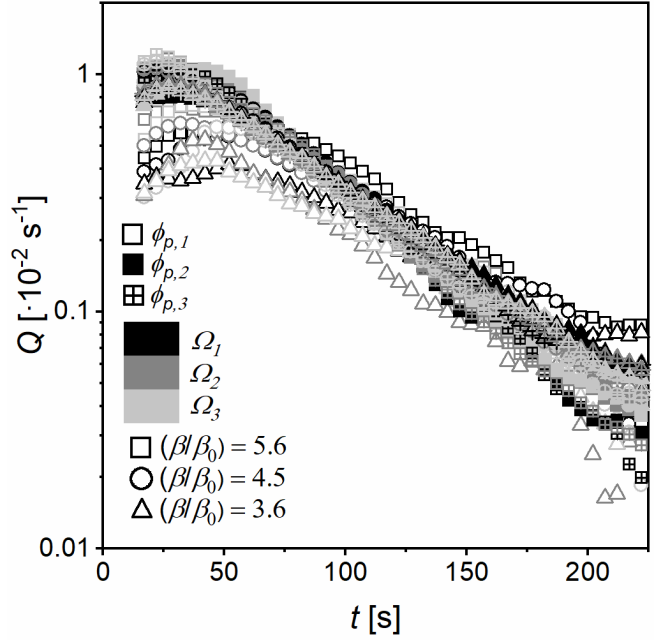

**Fig. 5. Particle flux through porous media.** A log-linear plot of the particle flux, $Q$, as a function of time for the same data shown in Figure 2. The symbol legend follows: squares, $(\beta/\beta_0) = 5.6$; circles, $(\beta/\beta_0) = 4.5$; triangles, $(\beta/\beta_0) = 3.6$; black, $\Omega_1 = 1.6$ Hz; dark gray, $\Omega_2 = 3.2$ Hz; light gray, $\Omega_3 = 4.8$ Hz; open, $\phi_{p,1} = 1.7 \times 10^{-3}$; filled, $\phi_{p,2} = 3.3 \times 10^{-3}$; and crossed, $\phi_{p,3} = 5.0 \times 10^{-3}$.

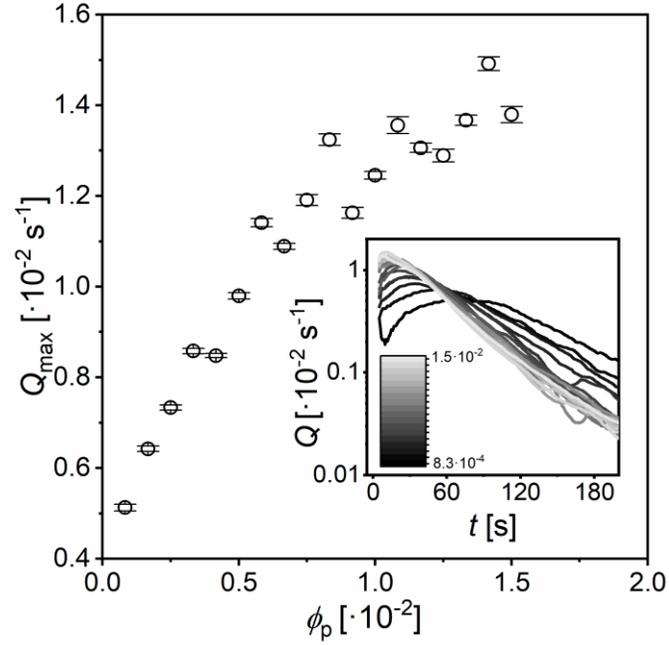

**Fig. 6. Influence of volume fraction on particle flux.** (**A**) Maximum flux values for increasing volume fraction of microrollers for $\phi_p = 8.3\times10^{-4} - 1.5\times10^{-2}$. Field strength and rotation rate are fixed at $(\beta/\beta_0) = 5.6$ and $\Omega = 3.2\ Hz$, respectively (**B**) Inset shows $Q$ for each $\phi_p$, where the maxima is taken between $c/c_{max} = 0.8$ and $0.4$ and $t$ is offset to zero. Error bars taken from confidence interval of the fit.

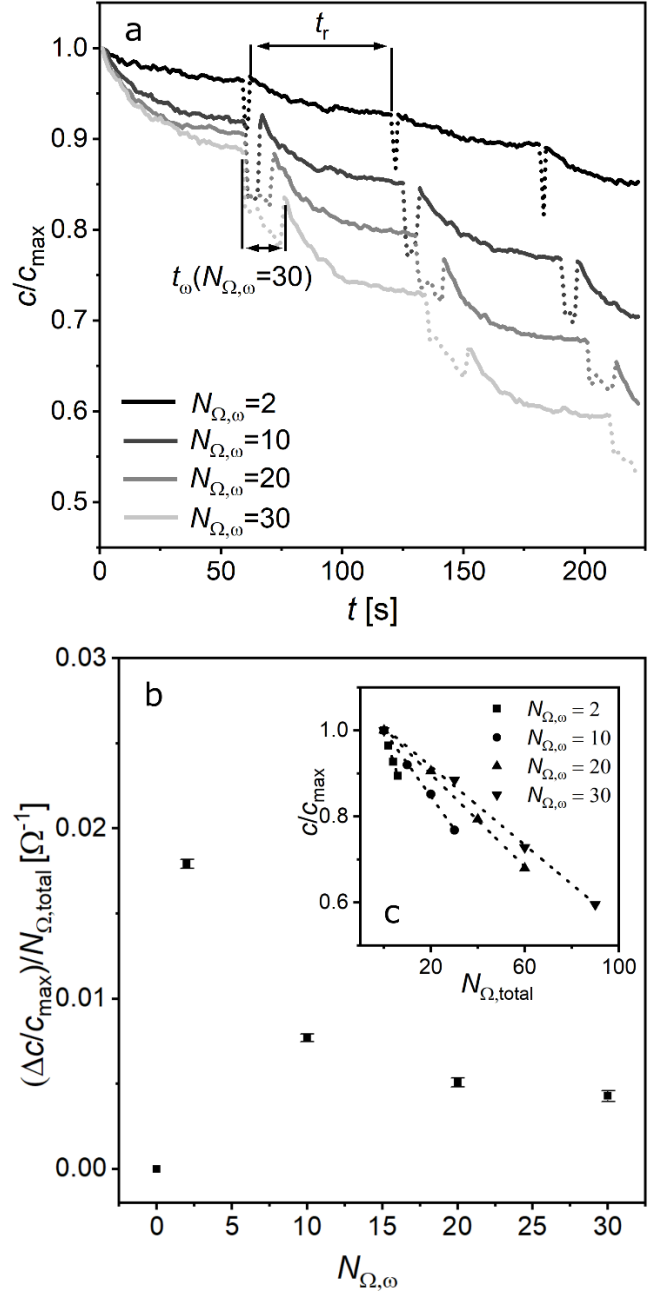

**Fig. 7. Energy cost to transport microrollers through porous media.** (**A**) Multi-cycle experiment for different Number of particle rotations per agitation cycle, $N_{\Omega,\omega}$. $t_r$ is the relaxation time between agitation cycles ( = 60 s) and $t_\omega$ is the sample dependent agitation time where $t_\omega \propto N_{\Omega,\omega}$. (**B**) Change in particle concentration $\Delta c/c_{max}$ in the network per total particle rotations, $N_{\Omega,total}$ against $N_{\Omega,\omega}$. (**C**) Cumulative $c/c_{max}$ with increased number of particle rotations, $N_{\Omega,total}$, for each agitation cycle length, $N_{\Omega,\omega}$. Data for $\phi_p = 1.7 \times 10^{-3}$ of microrollers in sample, $(\beta/\beta_0)=$ 5.6, $\Omega_2$.

## Supporting Information

### Materials

38-53 μm (measured; d[3,2] 43.6 ± 5.9 μm) polymethyl methacrylate microparticles were purchased from syringia lab supplies. Sodium chloride was purchased from Sigma-Aldrich. Hydrogel beads were purchased from eBoot.

### Equipment

4x 1000 G neodymium magnets (60 x 10 x 3 mm) were purchased from MIKEDE. Magnetic rotation was controlled with a programmable Lego-EV3 and motor. Magnetic field strength was measured with a Senjie SJ200 Teslameter, accuracy 1%, precision 0.1 mT. Back-lighting was provided by a Raleno 192 Video Light; 19.5 W. Containers were purchased from the Container Store.

**V1.** Background subtracted, macroscopic video of particles sedimenting within static bead network, followed by magnetic torque facilitated filtration.

**V2.** Microscopy video of microroller suspension sedimenting on gel beads followed by magnetic agitation with a rotating magnet positioned above the sample. Movie plays at 2 x speed.

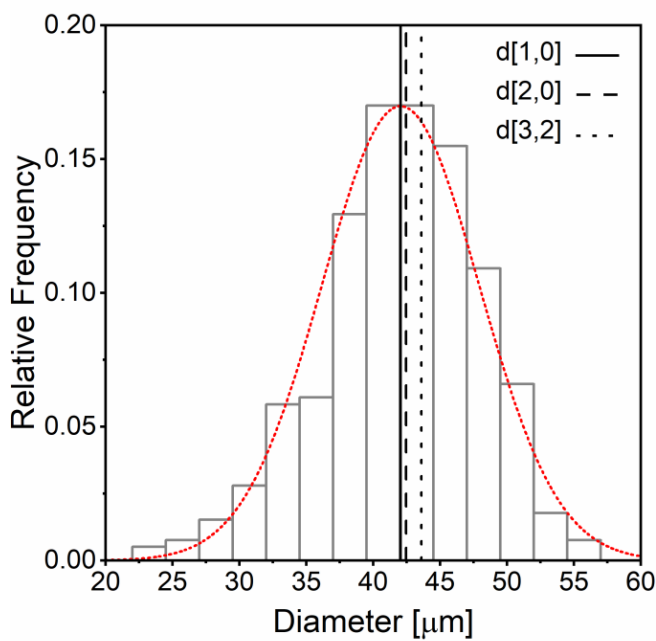

**Fig. S1** Particle size distribution measured by optical microscopy over 600 particles. Graph and data reused from previous publication using same preparation and batch of particles (OA License CC BY 4.O, https://creativecommons.org/licenses/by/4.0/).[33]

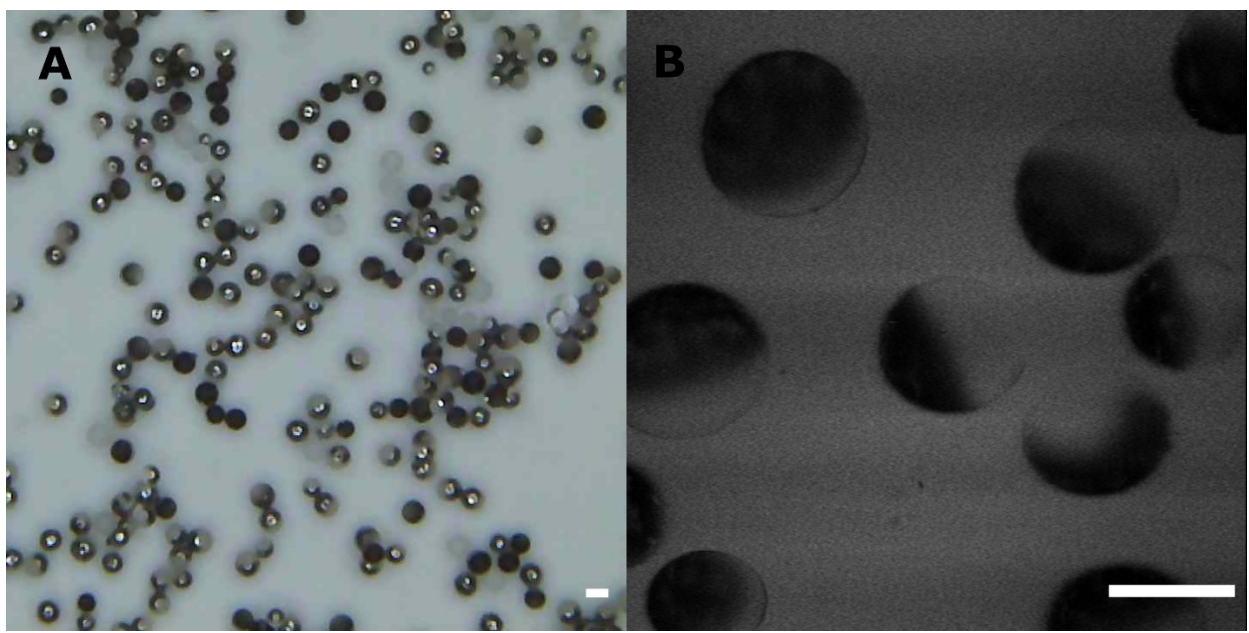

**Fig. S2** Microscopy of PMMA ferromagnetic Janus particle (a) Digital microscope image (b) confocal laser scanning microscopy. Scale bars = 43.6 µm (d[3,2]). Data resused from previous publication using same preparation and batch of particles (OA License CC BY 4.O, https://creativecommons.org/licenses/by/4.0/).[33]

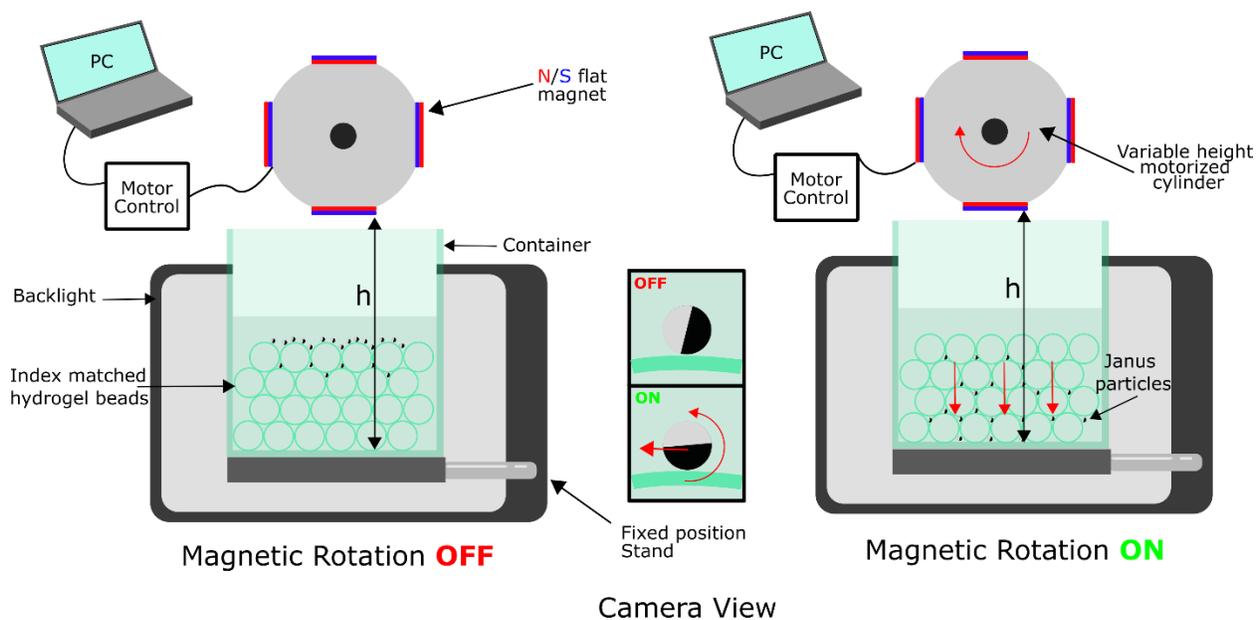

**Fig. S3** Diagram of experimental set-up in the driven "ON" and static "OFF" modes.

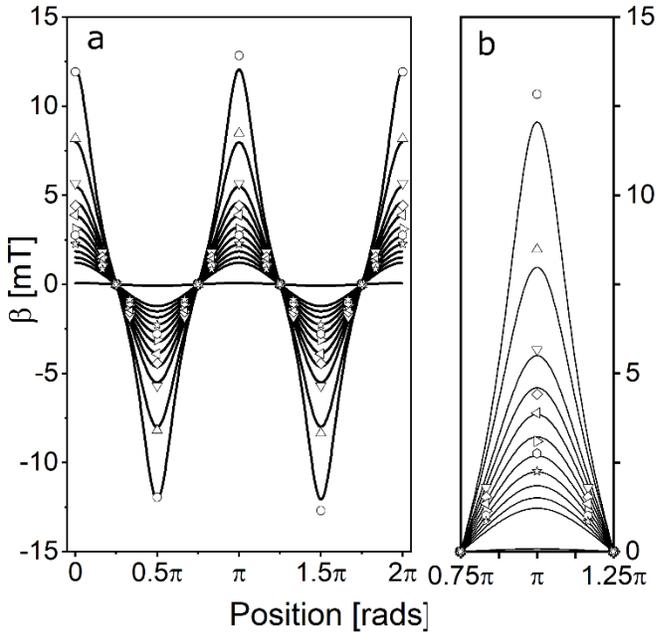

**Fig. S4** Fit of experimental field strength from 0-2π. Experimental data covers distances from the magnetic surface of $h$ = 15, 20, 25, 27.5, 30, 32.5, 35 and 37.5 mm. Graph and data resused from previous publication using same experimental aparatus (OA License CC BY 4.O, https://creativecommons.org/licenses/by/4.0/).[33]

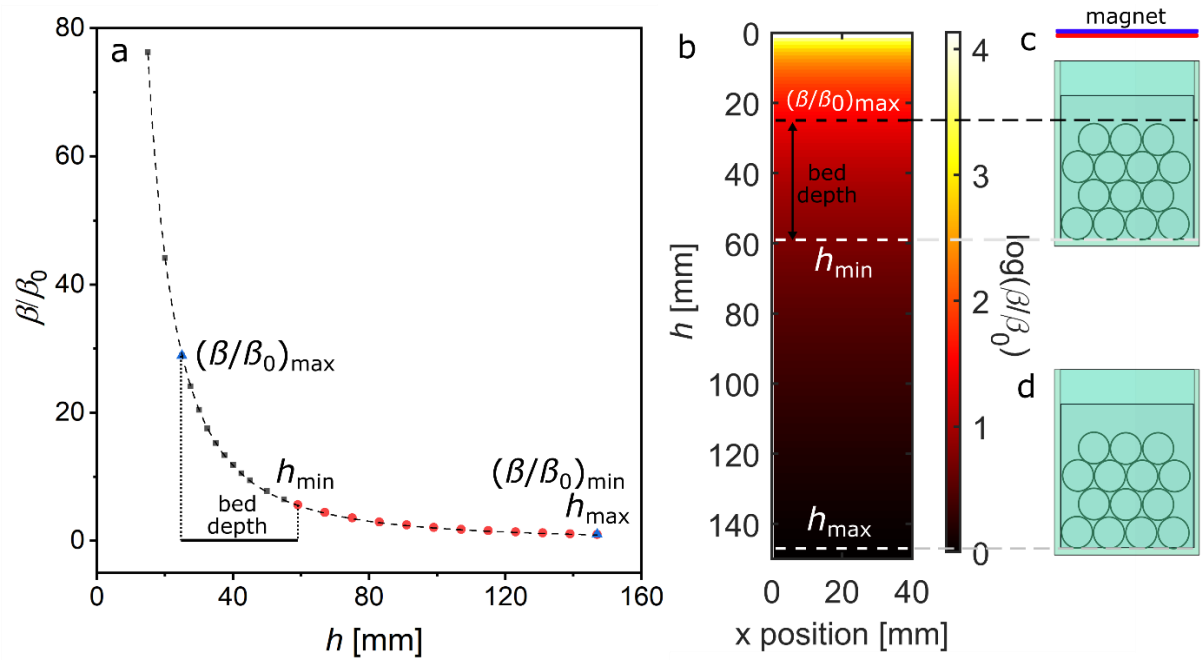

**Fig. S5** (a) Relationship between $\beta$ and height between magnet and sample bottom, $h$. (black squares) Full range of $\beta/\beta_0$ from 15 mm to 147 mm ($h_{max}$). (red circles) Experimental range of $\beta/\beta_0$ measured at the bottom of the samples. (Blue triangle) $(\beta/\beta_0)_{max}$ measured at the top of the bead packing at $h_{min}$ and $(\beta/\beta_0)_{min}$ measured at the bottom of the bead packing at $h_{max}$. (b) colormap of $\log(\beta/\beta_0)$ where y = 0 refers to magnetic surface and x position wis the width of the container (white dashes) show limits of $(\beta/\beta_0)$ at bottom of bead packing (black dash) shows $(\beta/\beta_0)_{max}$ at $h_{min}$ at the top of bead packing. (c-d) relative position of bead packings with respect to magnetic field and magnet position for $h_{min}$ and $h_{max}$.

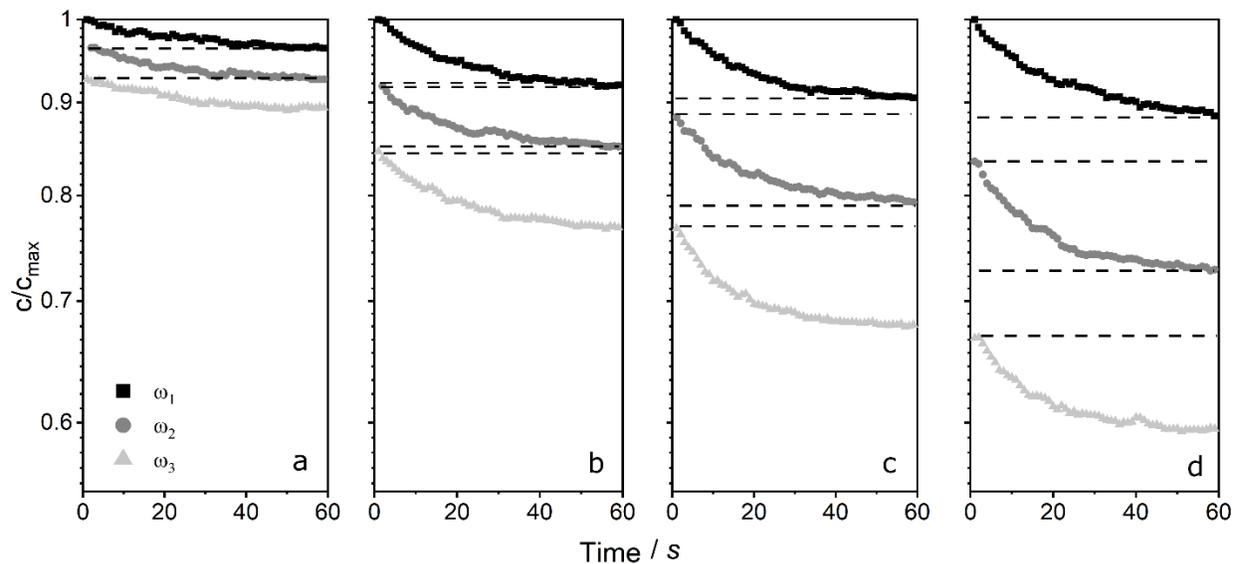

**Fig. S6** Curves related to data from Figure 5 of the main manuscript. $\omega_1$-$\omega_3$ indicate $\Delta c/c_{max}$ following successive agitation cycles. Samples are agitated for a number of rotations and then left to relax for 60 s. $\Delta c/c_{max}$ for the relaxation period is shown here. (a) $N_{\Omega,\omega} = 2$ (b) $N_{\Omega,\omega} = 10$ (c) $N_{\Omega,\omega} = 20$ (d) $N_{\Omega,\omega} = 30$. Dashed line regions indicate $\Delta c/c_{max}$ during the period where particle rotation is activated.